%% file: main.tex
\documentclass{eptcs}
\usepackage{program}
\usepackage{multicol}
\usepackage{mathpartir}

\def\aset#1{\left\{{#1}\right\}}
\def\aseq#1{\left\langle{#1}\right\rangle}
\def\apar#1{\left({#1}\right)}

\providecommand{\event}{LFMTP 2010}

\title{Generating Bijections between HOAS and the Natural
  Numbers\thanks{Work supported in part by the National Science Foundation
(CCF-0702635).  The opinions expressed here are not necessarily those
of the National Science Foundation or the US Government.}}
\author{John Tang Boyland
\institute{College of Engineering and Applied Science\\
University of Wisconsin--Milwaukee \\
Milwaukee, Wisconsin USA}
\email{boyland@cs.uwm.edu}}
\def\titlerunning{Generating Bijections}
\def\authorrunning{J.T. Boyland}

\begin{document}
\maketitle

\begin{abstract}
A provably correct bijection between higher-order abstract syntax
(HOAS) and the natural numbers enables one to define a ``not equals''
relationship between terms and also to have an adequate encoding of sets of
terms, and maps from one term family to another.  Sets and maps are
useful in many situations and are preferably provided in a library of
some sort.  I have released a map and set library for use with Twelf
which can be used with any type for which a bijection to the natural
numbers exists.

Since creating such bijections is tedious and error-prone, I have
created a ``bijection generator'' that generates such bijections
automatically together with proofs of correctness, all in the context
of Twelf.
\end{abstract}

\input intro.tex

\input adequate.tex
\input map.tex
\input proof.tex
\input impl.tex
\input related.tex
\input conc.tex

\bibliography{main}

\appendix

\section{Getting the Tool}

The bijection generator is available as
\begin{quote}
\texttt{\url{http://www.cs.uwm.edu/faculty/boyland/papers/map-natural.tar}}
\end{quote}
In order to use it, it will be necessary to first install
Scala~2.7~\cite{odersky:08scala2.7} and my Twelf library\\
(\url{http://www.cs.uwm.edu/faculty/boyland/proof/index.html}).

\end{document}

%% file: intro.tex
\section{Introduction}

Higher-order abstract syntax (HOAS)~\cite{pfenning/elliott:88hoas} uses the functions of the
meta-logic to represent functions (and related constructors, such as
``let''). For example (in each case, I define the canonical identity
function \verb|id|):
%\begin{minipage}{3in}
%\end{minipage}\hfill
%\begin{minipage}{3in}
%\end{minipage}
\begin{quote}
\begin{multicols}{2}\noindent
Traditional Abstract Syntax:
\begin{program}
t : type.
var : name -> t.
lam : name -> t -> t.
app : t -> t -> t.

\%abbrev id : t = lam x (var x).
\end{program}
Higher-Order Abstract Syntax:
\begin{program}
t : type.

lam : (t -> t) -> t.
app : t -> t -> t.

\%abbrev id : t = lam ([x] x).
\end{program}
\end{multicols}
\end{quote}
Not only does the traditional syntax need a type for ``names'' 
(where ``\verb|x|'' is a typical instance) but
then also must handle the fact that a variable may be undeclared.
Furthermore, there are the problems of accidental name clashes and
alpha-equivalence: the two functions \verb|lam X (var X)| and
\verb|lam Y (var Y)| are different if the names \verb|X| and \verb|Y|
are different.  The encoding of functions using names has
both ``junk'' and ``duplicates.''

Higher-order syntax maintains alpha-equivalence directly but only makes
sense in a logic in which the function cannot perform case analysis on
its argument.  One does not want the abstract syntax of a function to
depend on the semantic value of a parameter at run-time!

A technique that avoids duplicates is to use nameless terms (de Bruijn
terms).  In order to avoid junk as well, one uses an ``index'' on the
types~\cite{bird/paterson:99nested,licata/harper:07indexed}:
\begin{multicols}{2}
\begin{program}
vart : nat -> type.\label{type:vart}

1 : vart (s N).
1+ : vart N -> vart (s N).

term : nat -> type.

var : vart N -> term N.
lam : term (s N) -> term N.
app : term N -> term N -> term N.

\%abbrev id : term z = lam (var 1).
\end{program}
\end{multicols}
Here \verb|vart N| is the type of all legal variables inside \verb|N|
lambda abstractions; it has \verb|N| elements.
When \verb|N| is zero (written \verb|z|),
this type is empty.
Similarly \verb|term N| is the type of terms inside \verb|N| lambda
abstractions.

When one reasons with ASTs, it will sometimes be convenient to
determine whether two trees are equal or unequal, such as when one
forms sets of ASTs for analysis purposes.  It is fairly
straightforward (if tedious) to define inequality for traditional or
nameless terms, but with HOAS, this task is made much more complex
because one has to reason about inequality of functions.
Furthermore, as I describe in the following section, it is useful to
have a bijective mapping from ASTs to the natural numbers. 
The definition of inequality is trivial given a bijection. 

This paper shows how such mappings can be defined in general, and
proved correct.  I describe a tool I have implemented that produces such
mappings and proofs automatically.

In the following section, I review the concept of ``adequacy'' and
show how the desire for adequate encodings of sets and maps motivates
the definition of bijections between arbitrary term languages and the
natural numbers.  Section~\ref{sec:map} shows how such mappings
can be defined, with particular attention to HOAS and indexed types.
Then Section~\ref{sec:proof} explains how proofs can be constructed
to prove the correctness of the bijection.
Section~\ref{sec:impl} describes the implementation of a tool that
generates the mapping and correctness proofs in Twelf.

% LocalWords:  HOAS de Bruijn vart ASTs Twelf

%% file: adequate.tex
\section{An ``Adequate'' Encoding of (Finite) Sets}\label{sec:adequate}

Adequacy~\cite{harper/honsell/plotkin:93lf,harper/licata:07lf} is an
important concept in a
logical system---it means that
the source concepts (mathematical in nature) are faithfully
represented in the logical system.  In particular, an adequate
encoding is a bijective mapping from the source concept to a target
type for the logical
representation.  That is, the mapping is
\begin{description}
\item[total] Every instance of the source concept has an encoding.
\item[unique] The mapping is a function (``no confusion'').
\item[onto] Every instance of target type represents a valid
  instance of the source concept (``no junk'').
\item[one-to-one] Two distinguishable instances of source concept always
  map to distinguishable instances of the target type (``no loss'').
\end{description}
In this section, I demonstrate the concept of adequacy by showing how
it applies when encoding finite sets of natural numbers.

The first and last requirements (``total'' and ``one-to-one'') are
obviously needed for correctness.  If a set cannot be represented, or
if two different sets are represented the same way, the representation
obviously is not faithful.

The middle requirements (``unique'' and ``onto'') may also seem
obvious but are frequently violated in practice. For example, the
common practice of representing sets of natural numbers with lists of
natural numbers without duplicates violates both requirements:
the encoding is not unique because reorderings of a list
generates duplicate representations of the same set; neither is it onto
because lists of duplicates are ``junk'' that do not validly
represent any set.

What is the cost of having an encoding that is not unique and/or not onto?
One must define auxiliary relations.  In the case of a lack of
uniqueness, one must define an equivalence relation over the
representation type.  Furthermore, one must prove numerous lemmas that
say that equivalence is preserved by all operations on sets (union,
intersection, inclusion, containment, etc).  Since equivalence is
non-trivial when the encoding is not unique, these lemmas are also not
trivial.  Furthermore, equivalence ``invades'' all the uses of sets in
an application (i.e., a logical system that relies on sets): every
theorem concerning application relations touching on sets will need to
incorporate equivalence.

In the case of an encoding with junk, one must define a validity
relation, true for the subset of instances of the representation type
that have a source representation.  Such a validity relation is the
analog of a data structure invariant.  Again, it will be necessary to
prove that validity is preserved by set operations and by all
application relations using sets. If a set is in an ``input'' argument,
there must be an additional ``validity'' requirement on input.
Similarly, an ``output'' set value must be accompanied by a check of
validity.
If an encoding is neither unique nor onto, one will need both
equivalence and validity.

\begin{figure}[t]
\def\Map#1#2{\[\aset{#1} \Rightarrow \aseq{#2}\]}
\begin{multicols}{3}%
\Map{}{}
\Map{0}{0}
\Map{3}{3}
\Map{0,5}{0,4}
\Map{1,5}{1,3}
\Map{2,5}{2,2}
\Map{4,5}{4,0}
\Map{0,2,4}{0,1,1}
\Map{2,3,4}{2,0,0}
\Map{4,11,96}{4,6,84}
\end{multicols}
\caption{Illustration of an adequate encoding of finite sets of natural numbers.}
\label{fig:set-examples}
\end{figure}

Therefore, in order not to burden the application with equivalence
and/or validity, I defined an encoding of sets of natural numbers
that is fully adequate.\footnote{I expect others have done so as well,
  but there is apparently no accepted library of such signatures for
  Twelf.  My signature is released to the public domain.  This release
  includes other useful signatures, such as an adequate encoding of
  positive rational numbers.}
A set is represented by a sequence of numbers, the first of which is
the smallest element in the set and the remainder are the counts of
\emph{missing} numbers between adjacent (in sorted order) numbers in
the set.
Examples of this encoding are shown in
Figure~\ref{fig:set-examples}.

This technique for forming sets does not immediately apply to other
types: how would one form (say) a set of strings, of sets, or of HOAS
terms?  The difficulty is that one needs to be able to count the
elements ``between'' any two elements in the domain.  This requirement
is satisfied if one has a bijection from the element type to the
natural numbers.
Using the bijection, one can easily form sets of the element type by
mapping back and forth to integers and using the existing ``set''
signature.

Sets of course can be seen as a special case of maps, and indeed my
own set signature is created by specializing a map signature for the
``unit'' result type.  Thus a bijection between a type and the natural
numbers is useful for defining adequate encodings of (finite) mappings
as well.
This then is my motivation for exploring bijections between arbitrary
term languages and the natural numbers.

% LocalWords:  reorderings Twelf HOAS

%% file: map.tex
\section{Mapping Terms to Natural Numbers}\label{sec:map}

One of the delightful results of set theory is that the cardinality of
the set of pairs of natural numbers is the same as the cardinality of
the natural numbers themselves, in other words, that the set of pairs
of natural numbers can be arranged in a sequence with a
particular starting point.  A number of different simple mappings can
be defined.  I use the INTERCAL~\cite{woods/et.al:96intercal} ``mingle''
operation (written as infix ``\$''), which interleaves the bits of two
numbers to create a (unique) result:
{\def\0{\phantom{0}}
\[
\begin{array}{cccc}
0\$0 = 0 &

0\$1 = 1 &

0\$2 = 4 &

0\$3 = 5 \\

1\$0 = 2 &

1\$1 = 3 &

1\$2 = 6 &

1\$3 = 7 \\

2\$0 = 8 &

2\$1 = 9 &

\02\$2 = 12 &

\02\$3 = 13 \\

\03\$0 = 10 &

\03\$1 = 11 &

\03\$2 = 14 &

\03\$3 = 15
\end{array}
\]}
Figure~\ref{fig:nat} shows selected parts of the ``nat'' signature
used in this paper as well as some abbreviations used in the
generator.  The full definition of mingle is not shown.

\subsection{Simple Mappings}

\begin{figure}[tp]
\begin{minipage}[t]{3.0in}
\begin{verbatim}
nat : type.
z : nat.
s : nat -> nat.

plus : nat -> nat -> nat -> type.
plus/z : plus z Y Y.
plus/s : plus (s X) Y (s Z)
    <- plus X Y Z.

times : nat -> nat -> nat -> type.
times/z : times z X z.
times/s : times (s X) Y Z
    <- plus T Y Z

mingle : nat -> nat -> nat -> type.
...
\end{verbatim}
\end{minipage}
\begin{minipage}[t]{3in}
\begin{verbatim}
%abbrev 0 : nat = z.
%abbrev 1 : nat = s 0.
...

%abbrev 0+0=0: plus 0 0 0 = plus/z. 
%abbrev 1+0=1: plus 1 0 1 = plus/s 0+0=0.
%abbrev 0+1=1: plus 0 1 1 = plus/z.
...

%abbrev 0*0=0: times 0 0 0 = times/z.
...

%abbrev 0$0=0: mingle 0 0 0 = ... .
...
\end{verbatim}
\end{minipage}
\caption{Natural numbers, some operations and some abbreviations.}\label{fig:nat}
\end{figure}

A finite term type does not have a bijection to the natural numbers
(of course) but rather only to a subset of the same size.
finite types (such as booleans) are mapped to consecutive natural numbers
$0..n-1$.

When a term type has multiple constructors, then the
mapping I define distinguishes the
``finite'' constructors (whose instances are enumerated) from the
infinite ones.  For example:
\begin{multicols}{2}
\begin{program}
natlist : type.

natlist/0 : natlist.
natlist/+ : nat -> natlist -> natlist.
\end{program}

\begin{eqnarray*}
L(\texttt{natlist/0}) &=& 0 \\
L(\texttt{natlist/+ $n$ $l$}) &=& 1 + n\$L(l)
\end{eqnarray*}
\end{multicols}
When there are multiple ``infinite'' constructors, the mapping
results are distinguished through multiplication, as seen in the
following mapping for my rational type (based on continued fractions):

\medskip

\noindent
{\def\eq{&\!\!\!\!\smash{=}\!\!\!\!&}
\begin{minipage}{1.9in}
\begin{program}
rat : type.

whole : nat -> rat.
frac : nat -> rat -> rat.
\end{program}
\end{minipage}\hfill
\begin{minipage}{2.05in}
Bijection:\\[0.2ex]
\[
\begin{array}{rcl}
R(\texttt{whole $n$}) \eq  2n + 0\\[0.4ex]
R(\texttt{frac $n$ $r$}) \eq 2(n\$R(r)) + 1
\end{array}
\]
\end{minipage}\hfill
\begin{minipage}{2.05in}
Not a bijection:
\begin{eqnarray*}
R'(\texttt{whole $n$}) \eq 2n + 1\\
R'(\texttt{frac $n$ $r$}) \eq 2(n\$R'(r)) + 0
\end{eqnarray*}
\end{minipage}}

\medskip

Generating such mappings is fairly easy, although even with such
simple types, one must be careful: the superficially similar mapping
$R'$ is not onto. This fact can be seen since a $t$ for which
$R'(t) = 0$
would need to be of the form $t={}$\texttt{frac z $r$}, where
$R'(r)$ would again have to be zero.

Ignoring the latter issue, it is simply then a matter of
handling data type definitions with a mixture of finite and infinite
constructors, each of which may use subterms of finite type,
and whose infinite constructors may have any number of subterms of any
infinite type.  The situation is more interesting in the presence of
HOAS.

\subsection{Mapping functions}

With higher-order abstract syntax (HOAS), the mapping
essentially connects each variable with a small integer (in the range
$0..N-1$ where $N$ is the number of variables in scope).
For example, consider the pure lambda calculus defined previously:
\begin{multicols}{2}
\begin{program}
t : type.

lam : (t -> t) -> t.
app : t -> t -> t.
\end{program}
\begin{eqnarray*}
M_N(\texttt{lam $\lambda x$.$t$}) &=& N + 2\apar{M_{N+1}(t)} + 0 \\
&& \textrm{ where }  M_X(x) = N\\
M_N(\texttt{app $t_1$ $t_2$}) &=& N + 2\apar{M_N(t_1) \$ M_N(t_2)} + 1
\end{eqnarray*}
\end{multicols}
The extra parameter $N$ on the mapping indicates the number of free
variables in scope.  One uses $M_0(t)$ to determine the mapping for a
fully-bound term $t$.
Here the definition of the mapping uses ``where'' clauses to express
``hypothetical judgments.''  There is no
``\texttt{var}'' case since there is no ``\texttt{var}'' constructor
for the type.  The alert reader will notice that the mapping of a
variable does not change when it is used inside nested lambda
abstractions; it uses what Pierce~\cite{pierce:02types} calls ``de
Bruijn'' levels, not ``de Bruijn'' indices.

The nice thing about the mapping is that it preserves
alpha-equivalence---two terms at the same level map to the same
integer if and only if
they are the same (under alpha equivalence).  A consequence of the
bijection is that it is easy to define ``not equal'' over higher-order
terms via the mapping; such a relation is much harder to define
directly.

Here the example uses variables of an infinite type; one can also
define higher-order terms over finite types, but I have not seen an
application of this.  Therefore because it complicates the framework
(the cardinality of the finite type changes depending on the context),
my implementation does not handle variables of finite type.

\subsection{Indexed Types}

An indexed type~\cite{licata/harper:07indexed} uses one term to
distinguish different kinds of a second term.  A simple
example\footnote{This term family is a simplification of a typed predicate
system alluded to later.}
that
uses variables and indexing is the following
first-order functional language with recursion and a single constant:
\begin{program}
term : nat -> type.

unit : term z.
lam : (term z -> term N) -> term (s N).
app : term (s N) -> term z -> term N.
rec : (term N -> term N) -> term N.
\end{program}\label{loc:term}
Here \texttt{term $N$} is the type of terms that require $N$
arguments before they can be reduced to (first-order) values.
There is one canonical value, \verb|unit|; \verb|lam|
takes a function that accepts a value and returns a term requiring $N$
arguments and constructs one that now needs $N+1$ arguments;
\verb|rec| is used to create
a recursive function of $N$ arguments.

Now 
the mapping $M_V^n$ is qualified both by a ``vector'' (finite map of
nats to nats) $V$ and by $n$, the
index of the term.  The vector keeps track of how many variables of
the given index are currently free.
Each index has its \emph{own} bijection to the natural
numbers, with differing sets of constructors: \texttt{unit} for
the $n=0$ case only and \texttt{lam} for the $n>0$ case only.  
Thus for the $n=0$ case, the mapping handles
one finite constructor (\verb|unit|) and two infinite constructors
(\verb|app| and \verb|rec|), but the mapping for non-zero indices
($n+1$) does not handle any finite constructors and has three infinite
constructors:
\begin{eqnarray*}
M^0_V(\texttt{unit}) &=& V(0) \\
M^{0}_V(\texttt{app $t_1$ $t_2$}) &=& V(0) + 1+2\apar{M^{1}_V(t_1) \$ M^{0}_V(t_2)} + 0\\
M^{0}_V(\texttt{rec $\lambda f$.$t$}) &=& V(0)+1 + 2\apar{M^{0}_{V+0}(t)} + 1
\quad \textrm{ where }  M^{0}_X(f) = V(0)\\[1ex]
M^{n+1}_V(\texttt{lam $\lambda x$.$t$}) &=& V(n+1) + 3\apar{M^{n}_{V+0}(t)} + 0
\quad \textrm{ where }  M^0_X(x) = V(0)\\
M^{n+1}_V(\texttt{app $t_1$ $t_2$}) &=& V(n+1) + 3\apar{M^{n+2}_V(t_1) \$ M^{0}_V(t_2)} + 1\\
M^{n+1}_V(\texttt{rec $\lambda f$.$t$}) &=& V(n+1) + 3\apar{M^{n+1}_{V+(n+1)}(t)} + 2 
\quad \textrm{ where }  M^{n+1}_X(f) = V(n+1)\\
\end{eqnarray*}
Here $V(n)$ applies the vector to $n$ and
$V+n=V[n \mapsto V(n)+1]$ returns the vector with one more $n$.

\subsection{Binding variables}

Originally, I defined mappings using hypothetical judgments as
suggested earlier in this section.  However, separating the variable
information into a vector passed explicitly and hypothetical
judgments passed implicitly made the proofs of ``one-to-one'' (in
particular) extremely difficult.\footnote{%
The Twelf wiki shows the extreme lengths I went to for a small indexed
higher-order term system.}
Thus I changed the way variables were handled: the variables are now
passed explicitly in a large sequence that is then split in order to
find the location of a variable in the sequence.
This approach can be seen as another application of the idea of
``explicit contexts''~\cite{crary:explicit}.

\begin{figure}[tp]
\begin{verbatim}
var : term K -> type.
level : term K -> nat -> type.
nolevel : term K -> type.

%block block#var     : some {k}     block {x:term k} {v:var x}.
%block block#level   : some {k} {l} block {x:term k} {v:var x} {vl:level x l}.
%block block#nolevel : some {k}     block {x:term k} {v:var x} {nl:nolevel x}.
\end{verbatim}
\caption{Uninhabited types used in hypothetical judgments.}\label{fig:uninhab} 
\end{figure}

\begin{figure}[tp]
\begin{verbatim}
list : type.

list/0 : list.
list/+ : {k:nat} {x:term k} (var x) -> list -> list.
\end{verbatim}
\caption{Definition of variable list type.}\label{fig:list}
\end{figure}

In the end, I needed \emph{both} approaches: one in which
a variable is bound with its ``level'' in the implicit context (the normal
Twelf technique) and one in which the level is implicit in the
location of the variable in the explicit context list (so called
``nolevel'').  Then I generate proofs that a use
of a variable with a level can be converted into one without and vice
versa.

Some of proofs require that a variable
is either in the explicit context without a level or in the implicit
context with a level but not both.  Other proofs require that
variables are distinguished from normal terms.  Thus the generator 
defines three
judgments only used hypothetically: \verb|var|, \verb|level| and
\verb|nolevel| (see Fig.~\ref{fig:uninhab}).

\begin{figure}[tp]
\begin{verbatim}
split : list -> term K -> list -> list -> type.
count : nat -> list -> nat -> type.

split/here : split (list/+ _ X _ L) X list/0 L.
split/there : split L X L1 L2 -> split (list/+ _ _ V L) X (list/+ _ _ V L1) L2.

count/0 : count _ list/0 z.
count/= : count K L N -> count K (list/+ K _ _ L) (s N).
count/!= : count K L N -> nat`ne K' K -> count K (list/+ K' _ _ L) N.
\end{verbatim}
\caption{Operations on variable lists.}\label{fig:list-ops}
\end{figure}

The two operations on lists are given in Fig.~\ref{fig:list-ops}.
The \verb|split| operation locates a variable within a list, and
\verb|count| determines its level, by counting how many variables (of
its type) are in the context before it.  Here \verb|nat`ne| is the
inequality operator for natural numbers.

\begin{figure}[tp]
\begin{verbatim}
map : {K:nat} list -> term K -> nat -> type.

map/level : level X N -> map K _ X N.
map/nolevel : nolevel X -> split H X _ L -> count K L N -> map K H X N.
map/unit : count z H NV -> map z H unit NV.

map/app : map (s z) H A0 N1 -> map z H A1 NA1 -> mingle N1 NA1 N2 -> 
    count z H NV -> times 2 N2 TN -> plus 0 TN PN -> plus 1 NV NA -> 
    plus PN NA N -> map z H (app A0 A1) N.

map/rec : ({x} {v} nolevel x -> map z (list/+ z x v H) (A0 x) N1) -> 
    count z H NV -> times 2 N1 TN -> plus 1 TN PN -> plus 1 NV NA -> 
    plus PN NA N -> map z H (rec A0) N.

map/lam : ({x} {v} nolevel x -> map K (list/+ z x v H) (A0 x) N1) -> 
    count (s K) H NV -> times 3 N1 TN -> plus 0 TN PN -> plus 0 NV NA -> 
    plus PN NA N -> map (s K) H (lam A0) N.

map/app1 : map (s (s K)) H A0 N1 -> map z H A1 NA1 -> mingle N1 NA1 N2 -> 
    count (s K) H NV -> times 3 N2 TN -> plus 1 TN PN -> plus 0 NV NA -> 
    plus PN NA N -> map (s K) H (app A0 A1) N.

map/rec1 : ({x} {v} nolevel x -> map (s K) (list/+ (s K) x v H) (A0 x) N1) -> 
    count (s K) H NV -> times 3 N1 TN -> plus 2 TN PN -> plus 0 NV NA -> 
    plus PN NA N -> map (s K) H (rec A0) N.
\end{verbatim}
\caption{Mapping for the example indexed type.}\label{fig:map}
\end{figure}

Finally, Fig.~\ref{fig:map} gives the mapping as defined by the
generator (cleaned up for readability).  The basic structure is the
same as seen for the $M$ definition earlier except that it uses
variable lists; just as with $M$,
there are separate cases for the zero and non-zero term types.
The \verb|map/level| case is used internally in
proofs as described in the following section.

% LocalWords:  INTERCAL cccc nat natlist rcl frac subterms HOAS de Bruijn nats
% LocalWords:  Twelf wiki nolevel ne

%% file: proof.tex
\section{Proving Correctness}\label{sec:proof}

In order for the bijection to be used in a proof system, one needs a
proof that the mapping is indeed a bijection.  To wit, one needs four
proofs---one of each of the four aspects of a bijection outlined in
Section~\ref{sec:adequate}.  This section describes how such a proof
can be constructed in Twelf.

\subsection{Simple Mappings}

Ignoring
HOAS and indexed terms, three of the theorems
(technically ``meta-theorems'' in Twelf) are straightforward:
totality is proved as a simple ``effectiveness'' lemma, uniqueness is
proved using the uniqueness of the operations (addition,
multiplication, mingle) that are used to define it and ``one-to-one''
is proved using the uniqueness of remainder arithmetic (to distinguish
the infinite cases).  The last theorem (one-to-one) is long
for complex types with many constructors, because one has to check
every case against every other case, but each conflicting case is
simply a matter of using arithmetic to produce a contradiction.
My ``nat'' signature provides all the necessary theorems on addition,
multiplication and division.

With the proof of ``onto,'' there remains a termination issue:
the theorem is demonstrating that given a natural number, one can
produce a term that maps to it.  This theorem is of course recursive
(inductive) in its natural number argument.  Thus in a recursive
(inductive) call to the theorem, the number must
decrease.  The result of the mingle operation is greater than
its inputs in all cases except when the result is 0 or 1.
Thus, the main cases of the theorem may be restricted to only work for
$n > 1$, and then to have special cases for $n=0,1$.
Even these cases may cause unfounded recursion (induction) if the
mapping was defined incorrectly (as in the case for $R'$ shown
earlier).

The proof of ``one-to-one'' is the proof that two terms
mapping to the same number are the same term. Since equality of terms
is defined in terms of identity, the generator needs to define a lemma for each
constructor that says that if all the pieces are equal then  the
result is equal too.  Without variables, these lemmas are all trivial
to prove.
%, for example:
%\begin{quote}
%\begin{verbatim}
%%theorem app-respects-t#eq:
%    forall* {A0:t} {B0:t} {A1:t} {B1:t}
%    forall
%        {EQ0:t#eq A0 B0}
%        {EQ1:t#eq A1 B1}
%    exists
%        {EQ2:t#eq (app A0 A1) (app B0 B1)}
%    true.
%
%- : app-respects-t#eq t#eq/ t#eq/ t#eq/.
%\end{verbatim}
%\end{quote}
%The last line shows the only case needed to prove this lemma.
Even with variables, the lemmas are not complex, but if the variable
type is indexed, there is an interesting complication 
(see Sec.~\ref{sec:proof:indexed}).

\subsection{Variables}

When proving totality, it makes a difference whether the
source is a regular term or
a variable (hypothetical in the context).
As anyone
who has used variables in Twelf knows, one cannot use normal case
analysis to distinguish variables.  Instead one needs to define an
auxiliary predicate that distinguishes them, and then an effectiveness
lemma that indicates that the case analysis is always possible.
Furthermore, the effectiveness lemma needs to be in its own context.

As mentioned in the previous section, the context is not used to define
the mapping for variables; rather an explicit list of 
variables is used to compute the mapping.  This technique
however is difficult to handle with the totality and uniqueness
theorems: how does one know that every variable is present at least once
in the list (totality) and no more than once (uniqueness).  Twelf's ``regular
worlds'' do not give a way to connect a relation's parameter (the list
of variables in this case) and the context.  Thus there would seem to
be an impasse.  

The original technique (in which I used hypothetical
judgments to bind each variable to its ``level,''  the number to which
it should be mapped) had none of these problems---totality and
uniqueness fall out
immediately.  
This suggested a solution: for totality, the proof would bind
the variable in the context to its level, and then after all its uses
had been taken care of, remove the level binding and replace it with
the technique of using the parameter list.    The lemma for replacing
a mapping dependent on a level in the context with one that uses the
variable list is defined as follows:\footnote{
I use the \texttt{\%theorem} syntax because (1) it distinguishes
meta-theorems from normal relations and (2) it is clearer for
non-Twelf experts.}

\pagebreak

\begin{quote}
\begin{verbatim}
%theorem map-remove-term-level:
    forall* 
        {K1} {K2}
        {FH:{x:term K1} (var x) -> list)}
        {F:term K1 -> term K2}
        {L:nat} {N:nat}
        {H1:list} {H2:list}
    forall
        {FM:{x:term K1} {v:var x} level x L -> map K2 (FH x v) (F x) N}
        {FS:{x:term K1} {v:var x} split (FH x v) x H1 H2}
        {C:count K1 H2 L}
    exists
        {FM':{x:term K1} {v:var x} nolevel x -> map K2 (FH x v) (F x) N}
    true.
\end{verbatim}
\end{quote}
Ignoring the implicit (\texttt{forall*}) arguments, the theorem has
three inputs (\texttt{FM}, \texttt{FS}, \texttt{C}) and one output
(\texttt{FM'}).  
Thus \verb|FM| says that the mapping can be computed
for the term \verb|(F x)| that may use \verb|x|
as long as there is a level defined for \verb|x|.  The inputs \verb|FS|
and \verb|C| say that \verb|x| is in the parameter list and
that the count yields the same value as the
level.  The result says the mapping does \emph{not} need the
level, it can be done merely using the list.
This lemma is straightforward to prove using induction: the base case
for changing a variable to use the context, and the inductive case
for \verb|lam| are shown:
\begin{verbatim}
- : map-remove-term-level ([x] [v] [l] map/level l) FS C
                          ([x] [v] [n] map/nolevel n (FS x v) C).

- : map-remove-term-level 
    ([x] [v] [l] map/lam ([x0] [v0] [n0] F x0 v0 n0 x v l) (FC x v) D3 D4 D5 D6) 
    FS C 
    ([x] [v] [n] map/lam ([x0] [v0] [n0] G x0 v0 n0 x v n) (FC x v) D3 D4 D5 D6)
    <- ({x0} {v0} {n0} map-remove-term-level 
         (F x0 v0 n0) 
         ([x] [v] split/there (FS x v)) C 
         (G x0 v0 n0)).
\end{verbatim}
This lemma (similar to Crary's \verb|cut| lemma for explicit
contexts~\cite{crary:explicit}) enables totality to be proved.

When proving uniqueness, there is already a mapping (indeed, two
mappings) using the list
(``no level'') technique, so the generator uses a lemma that puts the
levels back
before looking at the bodies of the mappings.  For this to work, it is
essential to use the fact that the variable occurs in only one place
in the list.  The generator essentially proves the ``reverse'' lemma for which
\verb|FM'| along with \verb|FS| and \verb|C| are inputs and
\verb|FM| is the output.
This is why the type of \verb|FS| carefully uses \verb|H1| and
\verb|H2| as the type of the suffix and prefix, so that they are known
to be independent of the variable, and thus not including it.
This permits one to ensure that a variable cannot have two places
within the list.

Variables don't add much to complicate the ``onto'' and ``one2one''
theorems, except for more special cases for mapping
onto 0 and 1.  

\subsection{Indexed Types}\label{sec:proof:indexed}

Indexed types complicate the
``one-to-one'' theorem, in particular the trivial lemmas that say
if one constructs a term with equal parts, the results are equal (identical).
The problem is that the lemmas cannot always be typed.
Suppose there were another constructor for our type
in which an index variable does not occur in the result type:
\begin{verbatim}
exists : {M:nat} (term M -> term N) -> term N.
\end{verbatim}\label{loc:exists}
Notice that \verb|M| is not used in the result type.
(The name \verb|exists| is used because if one defines an indexed predicate
syntax with existentials, the existential term will have this form.)
Then, the lemma that shows that equal parts produce an equal whole
would have the following definition:
\begin{quote}
\begin{program}
\%theorem exists-respects-eq:
    forall* \{A0\} \{B0\} \{C\}
        \{A1:term A0 -> term C\} \{B1:term B0 -> term C\}
    forall
        \{EQ0:nat`eq A0 B0\}
        \{EQ1:\{x : term \textrm{???}\} eq (A1 x) (B1 x)\}
    exists
        \{EQ2:eq (exists A0 A1) (exists B0 B1)\}
    true.
\end{program}
\end{quote}
The problem here is that the type of \verb|x| must be \verb|term A0| and
\verb|term B0| at the same time.  Of course, these types are the same
(as demonstrated by \verb|EQ0|) but the type system doesn't ``know''
this fact.  This conundrum is solved by defining an equality predicate
directly on the functions:
\begin{quote}
\begin{verbatim}
func-eq : (term K3 -> term K1) -> (term K4 -> term K2) -> type.
func-eq/ : func-eq F F.
\end{verbatim}
\end{quote}
This extra layer requires another lemma
in the mutual induction of lemmas in proving the result:
\begin{quote}
\begin{verbatim}
%theorem map-one2one/func-eq:
    forall*
        {K1:nat} {K'1:nat} {K2:nat} {K'2:nat} {H:list}
        {F1:term K1 -> term K2}
        {F2:term K'1 -> term K'2}
        {N1:nat} {N2:nat}
    forall
        {E1:nat`eq K1 K'1}
        {E2:nat`eq K2 K'2}
        {FM1:{x} {v} nolevel x -> map K2  (list/+ K1  x v H) (F1 x) N1}
        {FM2:{x} {v} nolevel x -> map K'2 (list/+ K'1 x v H) (F2 x) N2}
        {EN:nat`eq N1 N2}
    exists
        {EF:func-eq F1 F2}
    true.
\end{verbatim}
\end{quote}
Then \verb|exists-respect-eq| is changed to use
\verb|func-eq|: \verb|{EQ1: func-eq A1 B1}|.

The biggest problem however with handling indexed types is if they are not
``uniform.'' I call an indexed type 
\emph{uniform} if all instances of the type (for different indices)
have the same cardinality.  I stumbled on a practical example of a
non-uniform indexed type in typed arguments:
\begin{quote}
\begin{verbatim}
argtype : type.
argtype/0 : argtype.
argtype/+ : nat -> argtype -> argtype.

actuals : argtype -> type.
actuals/0 : actuals argtype/0.
actuals/+ : term K -> actuals A -> actuals (argtype/+ K A).
\end{verbatim}\label{loc:actuals/+}
\end{quote}
This type lets one ensure that in a call the formal types (not shown) and the
actual types match (``correct by construction'').
The \verb|actuals| indexed type is however not uniform because the
type \verb|actuals A| is sometimes finite (cardinality 1 in fact)
and sometimes infinite.  The indexed type \verb|vart| (defined on
page~\pageref{type:vart}) is also not uniform.

In the end, I decided that I would not try to support non-uniform
indexed types.  If one wishes to use the bijection generator, one
must alter the source type, for example by changing the constructor
to make it infinite:
\begin{quote}
\begin{verbatim}
actuals/0* : nat -> actuals argtype/0.
%abbrev actuals/0 = actuals/0* z.
\end{verbatim}
\end{quote}
Thus the problem is pushed back to the source type which now has
``junk'' in it (non-zero numbers in empty argument lists).

The counter-intuitive result is that my generator works on HOAS, but
not on the de Bruijn version of the same syntax.

\subsection{Complex Type Families}

In Twelf, there is no distinction between type families used to form
terms, and those used to form relations among terms, and indeed those
used to express meta-theorems about relations among terms.
What should a bijection generator do when presented with a type such
as the following?
\begin{quote}
\begin{verbatim}
plus : nat -> nat -> nat -> type.

plus/z : plus z N N.
plus/z : plus X Y Z -> plus (s X) Y (s Z).
\end{verbatim}
\end{quote}
Given the limitation just described, the answer is easy: ``nothing,
because \texttt{plus} is not uniform.''  Furthermore, as described in
the following section, my generator permits only a single index, not three
indices as seen here.

However, even looking beyond the current limitations of my tool, any
generator must be able to determine the cardinality of a type in order
to correctly define the mapping.  Determining the cardinality of a type is
obviously undecidable.  Thus at some point the tool must break down,
which is unsurprising.

% LocalWords:  Twelf HOAS nat Twelf's forall FS Crary's existentials EQ func eq
% LocalWords:  vart de Bruijn

%% file: impl.tex
\section{Implementation}\label{sec:impl}

In this section, I describe the tool that generates the mapping and
theorems for a source signature.  After outlining how to use it, I
describe its limitations and then to what extent it can be thought of
as ``correct.''

\subsection{Running the tool}

The generator reads in a signature \texttt{required.elf} that defines
natural numbers, the uninhabited \verb|void| type, and a boolean type.
It assumes the existence of all the relations and theorems from 
my publicly available \verb|nat.elf| and \verb|natpair.elf| signatures.

It then reads in the argument which should define only the terms for
which a bijection is desired.  Abbreviations are also permitted, and
expanded where they occur.  If it notices a problem, it halts
execution (or simply throws an exception \ldots).

It then generates a large signature on standard output: this defines
the mapping and the four theorems for each type.  Along the way it
defines many auxiliary relations and lemmas.
To avoid name clashes, it ensures that (except for the abbreviations
of the form shown in Fig.~\ref{fig:nat})
all named entities have the hash sign (\verb|#|)
somewhere in the name.

The generated signature for the simple indexed type on page~\pageref{loc:term}
has over 250 definitions in over 60KB of text.

\subsection{Power}

I wrote the tool in order to generate the bijection for my permission
system which has five sorts of terms, three of which are defined
together (permission, unit permission and formula) and two which are
independent of these (object, wrapping a nat; fraction, wrapping a
rational number) but all five of which are defined under a generalized
term type \verb|gterm|.  This type thus is indexed with a finite index
type.  Along with this, I define predicate formal and actual types
using indexing to ensure parameter matching.  Predicate calls are an
instance of a formula and hence a \verb|gterm|.  On the other
hand, predicate formal parameters can be any \verb|gterm|.
The predicate system has the equivalent of \verb|rec| (see above)
indexed on the argument types.  The generated signature has over 1000
definitions in over 300KB of text.

Thus I had a fairly large and complex system to test with.
For now, I only need the bijection for this system, but I wanted to be able
to add new kinds of permissions and be able to regenerate the mapping
and proofs automatically.
The system is not much tested beyond  this test case and has a number
of known limitations.

\subsection{Limitations}

As explained previously, no tool could do the task for all type
families because in general the task is undecidable.
But, the tool is fairly limited.  In particular:
\begin{itemize}
\item Variables can only be of infinite type.
 That is, any HOAS function must be over an infinite parameter type.
\item Indexed types must have only a single index.
\item Indexed types must be uniform.
\item The index type must not itself be indexed or be defined with
  HOAS.
\item Explicit abstraction \verb|{X:xxx} ...| must
be explicitly typed (\verb|xxx| must be specified) and 
is only permitted if
\verb|xxx| is an index type and \verb|X| does \emph{not} occur in the
result type of the constructor (e.g., \verb|exists| on
page~\pageref{loc:exists}).
\item Unbound variables are permitted only if the variable
  represents an index, the variable is
  \emph{not} attributed with a type and the variable is used in
  the result type of the constructor (e.g., \verb|actuals/+| on page~\pageref{loc:actuals/+}).
\item The result type of an indexed constructor can only use one level
  of pattern matching if the index is infinite. For example, a
  constructor for indexed type \verb|term : nat -> type| (the index,
  \verb|nat| is infinite) may have type \verb|... -> term (s N)| but not 
  \verb|... -> term (s (s N))| which would require multiple levels of
  pattern matching.
  If the index type is finite, no
  such restriction exists because the tool enumerates all instances
  of finite index types.
\end{itemize}

\subsection{Correctness, or rather incorrectness}

I am reasonably confident that the tool will work well for simple
small examples, but if the type family is complex, the tool may reject
it, die with a match failure or even get caught in an infinite loop.
Furthermore, the generated mappings and theorems may not type check and
even if so, the meta-theorems may fail their totality checks.

All these failings are acceptable because
the tool does not pretend to be in the
trusted core of a system.  If the generated mapping and theorems work,
then they are correct on their own.

% LocalWords:  nat natpair gterm HOAS

%% file: related.tex
\section{Related Work}\label{sec:related}

G\"odel numbers are used to represent formal terms as natural numbers
so that number theory can be applied to formal logic.
For this purpose the mapping need only be total, unique and
one-to-one; specific mappings I have seen used in logic are never
``onto'' and need not be.  Variables are usually handled using a fixed
(possibly infinite) set of distinct symbols.  There is no attempt
to maintain alpha-equivalence.

Unlike Twelf,
the Coq proof system~\cite{coq:8.2manual} has an extensive library.
It appears that the library includes two implementations of
finite sets: one using sorted lists and one using balanced binary trees.
Neither implementation is ``adequate'' in the sense described in
Section~\ref{sec:adequate}: sorted lists have an extra invariant that
is preserved by the operations, and the balanced binary tree technique
not only has an invariant but also a non-trivial equivalence relation.
I expect that the invariants and probably the equivalence relation are
handled through tactics defined for use with the type.

A reviewer observed that
the entire generator might be seen as something like a ``tactic'' in
Coq: a rule to help generate proofs.
This similarity is intriguing, but one large pragmatic difference is
that the tool is intended to be used to generate a signature (Twelf file)
that would then be used along with the normal hand-written signatures.
In Coq, the tactics are named in the proof scripts and are checked
whenever the proof needs to be checked.  My generator would
have to be much more robust and transparent were it to be used in this way.

As a system that generates proofs about code generated for a source
text, the tool described in this paper can be seen as a very simple
and limited certifying compiler~\cite{necula/lee:98cert}.

% LocalWords:  Twelf Coq

%% file: conc.tex
\section{Conclusions}

An adequate encoding for sets (and by extension, maps) motivates the
desire for bijections from arbitrary type families to the natural
numbers.  Mapping functions are defined by enumerating the instances of
``finite constructors'' and mapping them  to small numbers.  For the infinite
constructors, 
the mapping uses the INTERCAL
mingle operation, multiplication and addition to divide up the
remaining numbers.  HOAS variables are handled by reserving space
``under'' the finite constructors for variables.  Proofs are
generated for the four requirements of a bijection (total, unique,
onto and one-to-one).  Handling variables requires some Twelf-specific
idioms for handling contexts.  The ``onto'' proof requires lemmas to
handle mappings to 0 and 1 to base the induction on.

Whether a mapping even exists for a given type family is undecidable,
so the tool described includes a number of limitations, but it seems
to be powerful enough to handle many type families used in practical
syntaxes.

\subsection*{Acknowledgments}

I thank Rob Simmons for taking up my challenge to define a provable
bijection between HOAS and DeBruijn-based syntax.  His work helped me
understand much better how one can work with variables and contexts in
Twelf.  In particular, he showed how to define a \verb|case| construct,
with an effectiveness lemma in its own context and the requisite
``fake'' declarations to avoid falling afoul of auto-freezing.
I thank the anonymous reviewers who gave much constructive criticism,
some of which is now reflected in the paper.

\bigskip
\noindent
SDG

% LocalWords:  INTERCAL HOAS Twelf SDG

%% file: main.bbl
\begin{thebibliography}{10}
\providecommand{\bibitemstart}[1]{\bibitem{#1}}
\providecommand{\bibitemend}{}
\providecommand{\bibliographystart}{}
\providecommand{\bibliographyend}{}
\providecommand{\url}[1]{\texttt{#1}}
\providecommand{\urlprefix}{Available at }
\providecommand{\bibinfo}[2]{#2}
\bibliographystart

\bibitemstart{bird/paterson:99nested}
\bibinfo{author}{Richard~S. Bird} \& \bibinfo{author}{Ross Paterson}
  (\bibinfo{year}{1999}): \emph{\bibinfo{title}{De {Bruijn} Notation as a
  Nested Datatype}}.
\newblock {\sl \bibinfo{journal}{Journal of Functional Programming}}
  \bibinfo{volume}{9}(\bibinfo{number}{1}).
\bibitemend

\bibitemstart{crary:explicit}
\bibinfo{author}{Karl Crary} (\bibinfo{year}{2008}):
  \emph{\bibinfo{title}{Explicit Contexts in {LF}}}.
\newblock In: {\sl \bibinfo{booktitle}{3rd International Workshop on Logical
  Frameworks and Meta-languages: Theory and Practice}}.
\newblock \bibinfo{note}{Revised version at
  \url{www.cs.cmu.edu/~crary/papers/2009/excon-rev.pdf}}.
\bibitemend

\bibitemstart{harper/honsell/plotkin:93lf}
\bibinfo{author}{Robert Harper}, \bibinfo{author}{Furio Honsell} \&
  \bibinfo{author}{Gordon Plotkin} (\bibinfo{year}{1993}):
  \emph{\bibinfo{title}{A framework for defining logics}}.
\newblock {\sl \bibinfo{journal}{Journal of the {ACM}}}
  \bibinfo{volume}{40}(\bibinfo{number}{1}), pp. \bibinfo{pages}{143--184}.
\bibitemend

\bibitemstart{harper/licata:07lf}
\bibinfo{author}{Robert Harper} \& \bibinfo{author}{Daniel~R. Licata}
  (\bibinfo{year}{2007}): \emph{\bibinfo{title}{Mechanizing metatheory in a
  logical framework}}.
\newblock {\sl \bibinfo{journal}{J. Funct. Program.}}
  \bibinfo{volume}{17}(\bibinfo{number}{4-5}), pp. \bibinfo{pages}{613--673}.
\newblock \urlprefix\url{http://dx.doi.org/10.1017/S0956796807006430}.
\bibitemend

\bibitemstart{licata/harper:07indexed}
\bibinfo{author}{Daniel Licata} \& \bibinfo{author}{Robert Harper}
  (\bibinfo{year}{2007}).
\newblock \emph{\bibinfo{title}{An extensible theory of indexed types}}.
\newblock \bibinfo{howpublished}{Web published}.
\newblock \urlprefix\url{http://www.cs.cmu.edu/~rwh/papers/extidx/paper.pdf}.
\bibitemend

\bibitemstart{coq:8.2manual}
\bibinfo{author}{\mbox{The Coq development team}} (\bibinfo{year}{2008}):
  \emph{\bibinfo{title}{The Coq proof assistant reference manual}}.
\newblock \bibinfo{organization}{LogiCal Project}.
\newblock \urlprefix\url{http://coq.inria.fr}.
\newblock \bibinfo{note}{Version 8.2}.
\bibitemend

\bibitemstart{necula/lee:98cert}
\bibinfo{author}{George~C. Necula} \& \bibinfo{author}{Peter Lee}
  (\bibinfo{year}{1997}): \emph{\bibinfo{title}{The Design and Implementation
  of a Certifying Compiler}}.
\newblock In: {\sl \bibinfo{booktitle}{Proceedings of the ACM SIGPLAN '97
  Conference on Programming Language Design and Implementation}},
  ~\bibinfo{volume}{32}, \bibinfo{publisher}{ACM Press}, \bibinfo{address}{New
  York}, pp. \bibinfo{pages}{333--344}.
\bibitemend

\bibitemstart{odersky:08scala2.7}
\bibinfo{author}{Martin Odersky} (\bibinfo{year}{2008}).
\newblock \emph{\bibinfo{title}{The Scala Language Specification (Version
  2.7)}}.
\newblock \bibinfo{howpublished}{Web published}.
\newblock
  \urlprefix\url{http://www.scala-lang.org/docu/files/ScalaReference.pdf}.
\bibitemend

\bibitemstart{pfenning/elliott:88hoas}
\bibinfo{author}{Frank Pfenning} \& \bibinfo{author}{Conal Elliott}
  (\bibinfo{year}{1988}): \emph{\bibinfo{title}{Higher-order abstract syntax}}.
\newblock In: {\sl \bibinfo{booktitle}{Proceedings of the ACM SIGPLAN '88
  Conference on Prog ramming Language Design and Implementation}},
  ~\bibinfo{volume}{23}, \bibinfo{publisher}{ACM Press}, \bibinfo{address}{New
  York}, pp. \bibinfo{pages}{199--208}.
\bibitemend

\bibitemstart{pierce:02types}
\bibinfo{author}{Benjamin~C. Pierce} (\bibinfo{year}{2002}):
  \emph{\bibinfo{title}{Types and Programming Languages}}.
\newblock \bibinfo{publisher}{The {MIT} Press}, \bibinfo{address}{Cambridge,
  Massachussetts, USA and London, England}.
\bibitemend

\bibitemstart{woods/et.al:96intercal}
\bibinfo{author}{Donald~R. Woods}, \bibinfo{author}{James~M. Lyon},
  \bibinfo{author}{Louis Howell} \& \bibinfo{author}{Eric~S. Raymond}
  (\bibinfo{year}{1996}).
\newblock \emph{\bibinfo{title}{The INTERCAL Programming Language: Revised
  Reference Manual}}.
\newblock \bibinfo{howpublished}{Web published}.
\newblock \urlprefix\url{http://www.catb.org/~esr/intercal/intercal.ps.gz}.
\bibitemend

\bibliographyend
\end{thebibliography}
